# A theoretical investigation on the transport properties of armchair biphenylene nanoribbons

Hongyu Ge, Guo Wang*, Yi Liao

*Department of Chemistry, Capital Normal University, Beijing 100048, China*

\* Corresponding author. Tel.: +86-10-68902974.

*E-mail address:* wangguo@mail.cnu.edu.cn (G. Wang)

ABSTRACT

Armchair biphenylene nanoribbons are investigated by using density functional theory. The nanoribbon that contains one biphenylene subunit in a unit cell is a semiconductor with a direct band gap larger than 1 eV, while that containing four biphenylene subunits is a metal. The semiconducting nanoribbon has high electron mobility of 57174 $cm^2V^{-1}s^{-1}$, superior to armchair graphene nanoribbons. Negative differential resistance behavior is observed in two electronic devices composed of the semiconducting and metallic nanoribbons. The on/off ratios are in the order of $10^3$. All these indicate that armchair biphenylene nanoribbons are potential candidates for ultra-small logic devices.

## 1. Introduction

Graphene [1] has attracted great attention for its ultra-high carrier mobility [2]. For a logic device, a sufficiently large band gap is essential for on/off operation [3]. Unfortunately, graphene has an intrinsic band gap equal to zero [4], which makes it difficult to be switched off [1,2,5,6]. Although one-dimension quantum confinement opens its band gap [7,8], the lack of atomically precise edges of graphene nanoribbons fabricated by top-down methods significantly degrades the performance of graphene-based electronic devices [9]. A bottom-up approach [10] produces precise edges of graphene nanoribbon. However, only one type of armchair graphene nanoribbons with the width of seven carbon atoms can be synthesized through this method so far. These hinder their application in logic devices.

Beyond graphene, several carbon allotropes are also discovered, such as graphdiyne [11] and atomic carbon chain [12]. The era of carbon allotropes [13] is coming. Recently, the controlled functionalization of specific positions [14] of biphenylene [15] significantly promotes its polymerization and makes biphenylene nanoribbon a new candidate for carbon allotropes with excellent properties. Furthermore, theoretical investigations indicate that biphenylene nanostructures have appealing properties [16-19], such as low reorganization energies [18] and high capacity for hydrogen storage [19]. It is interesting that biphenylene nanoribbons are either semiconducting or metallic, depending on their width [16,17]. This facilitates the control of band gaps, which are crucial for electronic devices. Moreover, biphenylene nanoribbons are expected to be synthesized by bottom-up approaches [14], which could reduce edge roughness [9] and enhance device performance. It is noted that only the narrowest zigzag biphenylene nanoribbon is a semiconductor with a small band gap of 0.4 eV, while the band gaps of the armchair biphenylene nanoribbons (ABPNRs) can be much larger [16]. All these imply that ABPNRs can be good candidates for new carbon-based logic devices.

In the present work, ABPNRs are investigated by using density functional theory. The transport properties are focused on. The results indicate that the semiconducting ABPNR has high electron mobility. Furthermore, a negative differential resistance behavior is observed in the electronic devices composed solely of ABPNRs. The devices can be switched off with on/off ratios in the order of $10^3$.

## 2. Computational details

The geometries and electronic properties of the ABPNRs are calculated with the CRYSTAL14 program [20,21]. A pure density functional PBE as well as Bloch functions based on 6-21G(d,p) basis set are used. A band gap is an important parameter in electronic devices. However, pure density functionals usually underestimate the band gaps of solids. Therefore, a screened hybrid density functional

HSE06 [22], which can calculate band gaps accurately [23], is also used in the calculations. A Monkhorst-Pack sampling with 81 k-points in the first Brillouin zone is sufficient to obtain converged electronic properties. During the non-iterative band structure calculations, 801 k-points are used in order to fit the carrier effective masses accurately. For semiconductors without sharp density of states near their frontier band edges, carriers in an energy range that is wider than $k_BT$ could participate in conduction. The range 10 $k_BT$ [24] is used to fit the carrier effective masses. Under the deformation potential theory [25], the carrier mobilities of one-dimensional structures are calculated by [26]

$$\mu_{1D} = \frac{e\hbar^2 C_{1D}}{(2\pi k_B T)^{1/2} |m^*|^{3/2} E_1^2} \quad (1)$$

where $C_{1D} = a_0 \partial^2 E / \partial a^2 |_{a=a_0}$ is the one-dimensional stretching modulus, $a_0$ is the lattice constant at equilibrium geometry, $E$ is the total energy, $m^* = \hbar^2 [\partial^2 \varepsilon / \partial k^2]^{-1}$ is the carrier effective mass, $\varepsilon$ is the energy at the frontier band edge, $k$ is the reciprocal lattice vector, and $E_1 = a_0 \delta \varepsilon / \delta a |_{a=a_0}$ is the deformation potential constant. The deformation potential theory has been successfully applied to similar one-dimensional structures, such as graphene nanoribbons [27,28].

Based on the non-equilibrium Green's function method, the current-voltage (*I-V*) characteristics of the electronic devices composed of ABPNRs are calculated according to the Landauer-Büttiker formula [29]

$$I = \frac{2e}{h} \int_{-\infty}^{+\infty} T(E, V_b, V_g)[f_L(E) - f_R(E)] dE \quad (2)$$

in which $T(E,V_b,V_g)$ is the transmission coefficient at energy $E$, bias $V_b$ and gate voltage $V_g$, $f_L(E)$ and $f_R(E)$ are the Fermi-Dirac distribution function at the left and right electrodes, respectively. A density functional PBE and norm-conserving pseudopotentials in the OPENMX program [30] are used. Pseudoatomic orbitals with cutoff radii of 5.0 Bohr are adopted as basis functions, in which one primitive orbital is used for each of the s or p orbital. The size of vacumm layer is set to 15 Å. The energy cutoff is 150 Ry. A Monkhorst-Pack k-mesh 121×1×1 is sufficient to obtain converged properties.

## 3. Results and discussions

Four one-dimensional ABPNRs are calculated based on the PBE functional by using CRYSTAL14 program. The width is indicated by the number of biphenylene subunits in a unit cell, which is labeled before the abbreviation. For example, 1-ABPNR indicates that the unit cell contains only one biphenylene subunit. The structure of 1-ABPNR is indicated in Figure 1(c) and its band structures are shown in Figure 1(a). From Figure 1(a), it can be seen that 1-ABPNR is a semiconductor with a direct band gap of 1.13 eV at the Γ point. For 2-ABPNR, the valence band maximum (VBM) moves to the X point, while the conduction band minimum (CBM) is still at the Γ point. The indirect band gap is only 0.09 eV. The 3-ABPNR and 4-ABPNR are both metals. This is similar to the results obtained by the M06-L functional [17]. Because 3-ABPNR and 4-ABPNR have similar metallic band structures, only those of 4-ABPNR are shown in Figure 1(b). In the figure, the frontier bands go across the Fermi level three times. The density of states at the Fermi level is 4.1 eV$^{-1}$cell$^{-1}$. Since the HSE06 functional can give much more accurate band gaps of solids, the four ABPNRs are also calculated by this functional. The band gap of 1-ABPNR increases from 1.13 to 1.71 eV, while that of 2-ABPNR increases from 0.09 to 0.53 eV. 3-ABPNR is no longer a metal, but a semiconductor with a very small band gap of 0.08 eV. 4-ABPNR is still a metal. The results obtained by the HSE06 functional are consistent with the previous results [16,17]. Whatever the adopted functional is, 1-ABPNR is a semiconductor with a band gap larger than 1 eV while 4-ABPNR is a metal. 4-ABPNR can be used as an electrode, and 1-ABPNR can be used as a semiconductor material. Since the HSE06 functional is still not available in *I-V* characteristics calculations [30], all the results below are based on the PBE functional for consistency.

Carrier mobility is a key parameter in semiconductor industry. According to the deformation potential theory, carriers are mostly scattered by longitudinal acoustic phonons. The valence and conduction band deformation potential constants ($E_{1v}$ and $E_{1c}$) of the semiconducting 1-ABPNR are 8.86 and 0.49 eV, respectively. The $E_{1v}$ is more than one order of magnitude larger than the $E_{1c}$. The unbalanced constants can be explained using frontier crystal orbitals. In Figure 1(c) and 1(d), the highest occupied crystal orbital (HOCO) at the VBM and the lowest unoccupied crystal orbital (LUCO) at the CBM are both π orbitals. The HOCO is almost localized (perpendicular to the one-dimensional direction), while the LUCO is almost delocalized. During the deformation along the one-dimensional direction, the delocalized orbital should have a smaller energy change than the localized one has. Therefore, the $E_{1c}$ is smaller than the $E_{1v}$. Furthermore, there is little orbital distributed above the center of the slanted C-C bonds, because the unoccupied orbital has more nodes than the occupied one has. Due to the nodes at all the slanted C-C bonds, the $E_{1v}$ is extremely small. According to equation (1), carrier effective mass is another parameter that affects carrier mobility. Carrier effective masses are closely related to band dispersion. As shown in Figure 1(a), the band width of the valence band (0.26 eV) is less than that of the conduction band (0.98 eV). Thus the fitted hole effective mass (0.95 $m_0$) is larger than the electron effective mass (0.20 $m_0$). Besides, the calculated stretching modulus is 153 eVÅ$^{-1}$. The obtained hole and electron mobilities are 17 and 57174 cm$^2$V$^{-1}$s$^{-1}$, respectively. This indicates that 1-ABPNR is favorable to electron transport and could be a candidate for high-speed electronic devices. The high electron mobility is a result of the small $E_{1v}$ and small electron effective mass.

The armchair graphene nanoribbon with the width of seven carbon atoms (7-AGNR) is the only one of the AGNRs synthesized by a bottom-up approach [10]. It has the advantages of atomically smooth edges. The carrier mobilities are also calculated with the same method for comparison. The stretching modulus is 196 eVÅ$^{-1}$, which is 1.28 times as large as that of 1-ABPNR. It is noted that the width of 7-AGNR is larger than that of 1-ABPNR (six carbon atoms). Considering this difference, the stretching modulus should be 1.10 times as large as that of 1-ABPNR. The perfect honeycomb structure makes graphene nanoribbons stronger than any other counterparts. Similar to 1-ABPNR, 7-AGNR has a direct band gap at the Γ point. It has a slightly larger band gap of 1.52 eV. The $E_{1v}$ and $E_{1c}$ are 3.27 and 10.39 eV, while the hole and electron effective masses are 0.33 and 0.38 $m_0$, respectively. These make 7-AGNR be favorable to hole transport. The obtained hole and electron mobilities are 759 and 63 cm$^2$V$^{-1}$s$^{-1}$, respectively. The hole mobility is comparable to the value of much wider graphene nanoribbons obtained by top-down methods [9]. The electron mobility of 1-ABPNR (57174 cm$^2$V$^{-1}$s$^{-1}$) is more than one order of magnitude higher than the mobilities of 7-AGNR, and is of the same order of magnitude as that of much wider 39-AGNR [28]. All these imply that 1-ABPNR should be a good candidate for ultra-small electronic devices.

With the size of electronic devices becomes smaller and smaller, short-channel effect can occur. Low-dimensional structures in electronic devices should be helpful to solve this problem. Instead of carrier mobilities, real performance (*I-V* characteristics) is usually used to describe the properties of ultra-small electronic devices. Two all-ABPNR devices are proposed as shown in Figure 2(a) and 2(b). The semi-infinite left and right electrodes are both composed of metallic 4-ABPNR. The scattering region contains a unit cell of 4-ABPNR at each terminal as well as five cells of semiconducting 1-ABPNR. Metal electrodes are not used in the models in order to prevent possible contact barrier between metals and 1-ABPNR. In Figure 2(a) or 2(b), the 1-ABPNR is at the edge or in the center. In the "edge model", the hydrogen atoms at the inner edge of 1-ABPNR have repulsion with the adjacent hydrogen atoms of 4-ABPNR. This makes 1-ABPNR slightly deviate from the plane of 4-ABPNR. The largest deviation is 8° near the inner edge of 1-ABPNR. In the "center model", the repulsion exists at the both edges of 1-ABPNR. The deviation increases to 10°. The channel length is only 31 Å for the two ultra-small devices.

The *I-V* curves of the two models are shown in Figure 3(a) and 3(b). Since the left and right electrodes are symmetric, only positive bias is considered. The *I-V* curves of the two models are similar and they have semiconducting characteristics. The edge and center models have threshold voltages of 0.7 and 0.6 V, respectively. Below the thresholds, the currents are always close to zero. Above the thresholds, the currents increase rapidly with the biases. When the biases are equal to 2.0 V, the currents of the two models are both the largest. The largest currents are 30.5 and 17.0 µA. Then they begin to decrease until the biases reach 2.6 and 2.4 V. Negative differential resistance exists in these devices. The valley values are 13.6 and 10.3 µA. The peak-to-valley current ratios are 2.24 and 1.65 for the two models, respectively. The edge model has a larger peak current and a higher peak-to-valley current ratio. The performance is slightly better than that of the center model. The reason may be the smaller structural deviation and better junction in the edge model. It is noted that there is another valley at 2.8 V for the center model. The corresponding peak-to-valley current ratio is lower than the one at 2.4 V.

In order to elucidate the *I-V* characteristics, transmission spectra at different biases are plotted. For the edge model, the transmission coefficient at zero bias is shown in Figure 4(a). There is a gap between -0.7 and 0.5 eV, where the transmission coefficient is almost zero. Thus the current is almost zero at small biases. On the other hand, three peaks exist in the range from 0.5 to 1.5 eV. The above two characteristics are basically maintained when the bias is equal to 0.7 V, which can be seen from Figure 4(b). However, the peaks shift slightly to the left and the peak values decrease. For a bias $V_b$, the transport window is from $-eV_b/2$ to $eV_b/2$ when Fermi level is set to zero. The transport window is indicated by dotted vertical lines. In Figure 4(b), there is a small non-zero area in the transport window, so the integral in equation (2) is non-zero. Then the current increases with the bias, because more area is included in the transport window. When the bias is equal to 2.0 V, the middle one of the three peaks becomes dominant, which is shown in Figure 4(c). The biggest transmission peak is included in the transport window and the current is the largest. As shown in Figure 4(d), the heights of the three peaks drastically decrease, when the bias increases further to 2.6 V. The current decreases, although the three transmission peaks are all included in the transport window. This is the reason why negative differential resistance occurs. The situation for the center model is similar and is not shown for brevity.

For logic devices, on/off operation is an important issue. The speed of a field-effect transistor is proportion to conductance [31]. The conductance reaches the highest value when the bias is 2.0 or 1.5 V for the edge or center model, respectively. In order to avoid potential influence from the negative differential resistance effect above 2 V, the bias 1.5 V is chosen for on/off operation. At this bias, the conductance is 8.1 or 9.5 μS for the edge or center model, indicating considerably high performance. For the edge model, the current is 12.1, 0.369, 0.0101 or 0.0111 μA when the gate voltage $V_g$ is 0, 5, 10 or 15 V. Negative $V_g$ does not effectively switch the device off. For the center model, the current is 14.2, 0.108, 0.0411, 0.00734 or 0.0703 μA when the $V_g$ is 0, 5, 10, 15 or 20 V. The highest on/off ratio of the edge or center model is $1.2 \times 10^3$ and $1.9 \times 10^3$ when the $V_g$ is 10 or 15 V. The ratios are much higher than that of graphene [5,6], and are close to the requirement ($10^4$) of complementary circuits [3].

## 4. Conclusions

Four one-dimensional ABPNRs are investigated by using density functional theory. The calculations indicate that 1-ABPNR is a semiconductor with a direct band gap larger than 1 eV, while 4-ABPNR is a metal. The semiconducting 1-ABPNR has high electron mobility of 57174 $cm^2V^{-1}s^{-1}$, which is calculated based on the deformation potential theory. The high electron mobility is a result of the small $E_{1v}$ and small electron effective mass. This value is more than one order of magnitude higher than those of 7-AGNR. And 7-AGNR is the only one of the AGNRs with atomically smooth edges synthesized by bottom-up approaches. The electron mobility of 1-ABPNR is of the same order of magnitude as that of much wider 39-AGNR. These indicate that 1-ABPNR is a good candidate for ultra-small electronic devices with high speed. Two models of electronic devices composed of 4-ABPNR and 1-ABPNR are proposed. The *I-V* characteristics of the electronic devices are calculated based on the non-equilibrium Green's function method. The two electronic devices have threshold voltages of 0.7 and 0.6 V. Negative differential resistance occurs when the bias is larger than 2.0 V. This is explained using transmission spectra. The peak-to-valley current ratios are 2.24 and 1.65 for the two models, respectively. Furthermore, the devices can be switched off by applying gate voltages. The on/off ratios are in the order of $10^3$. These imply that ABPNRs are potential candidates for ultra-small logic devices.

## Acknowledgements

This work is supported by the National Natural Science Foundation of China (Grant No. 21203127), the Beijing Higher Education Young Elite Teacher Project (YETP1629) and the Scientific Research Base Development Program of the Beijing Municipal Commission of Education.

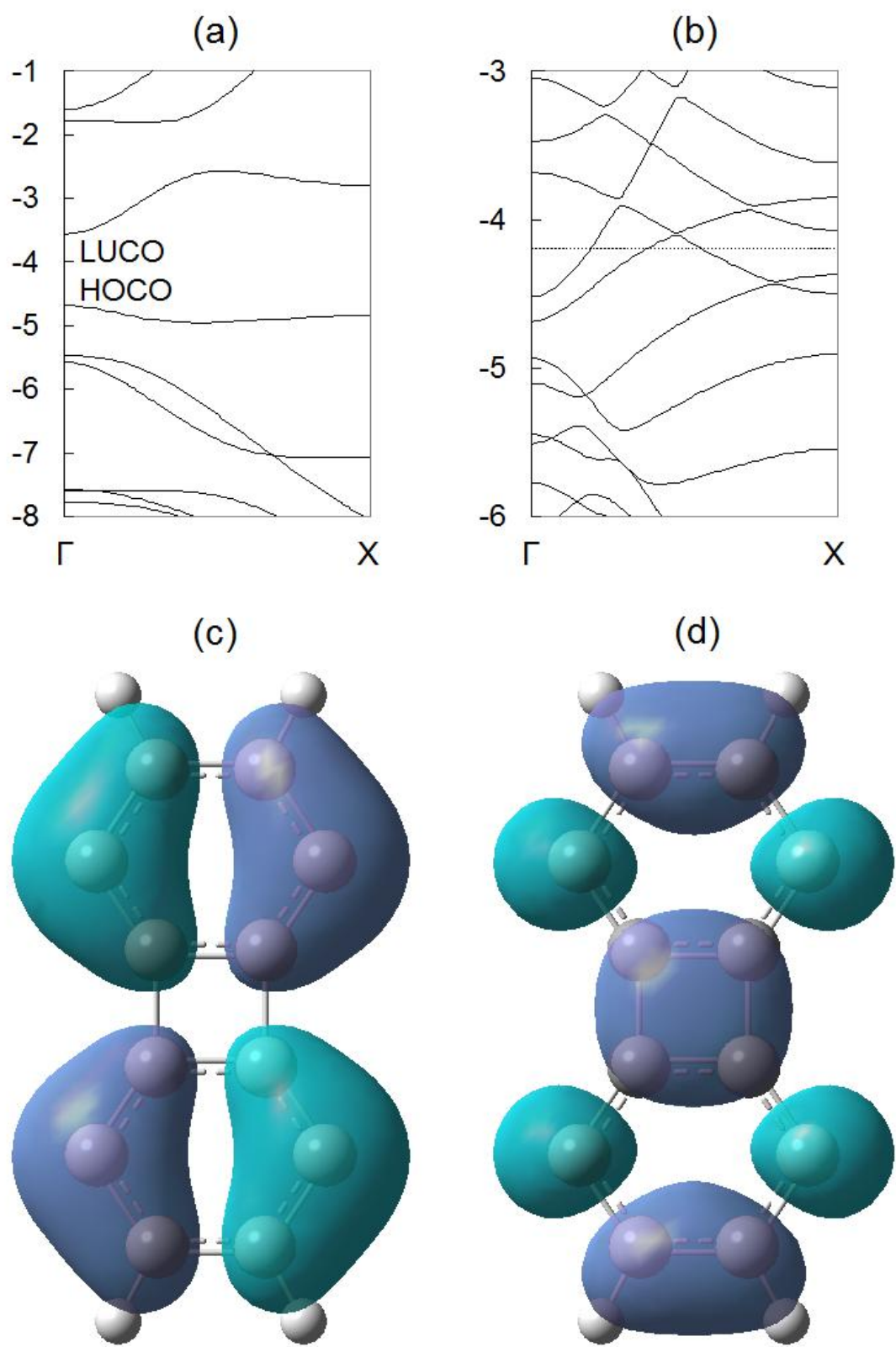

**Figure 1.** Band structures of (a) 1-ABPNR and (b) 4-ABPNR. Horizontal axis: reciprocal lattice vector, vertical axis: energy (eV). (c) HOCO and (d) LUCO of 1-ABPNR.

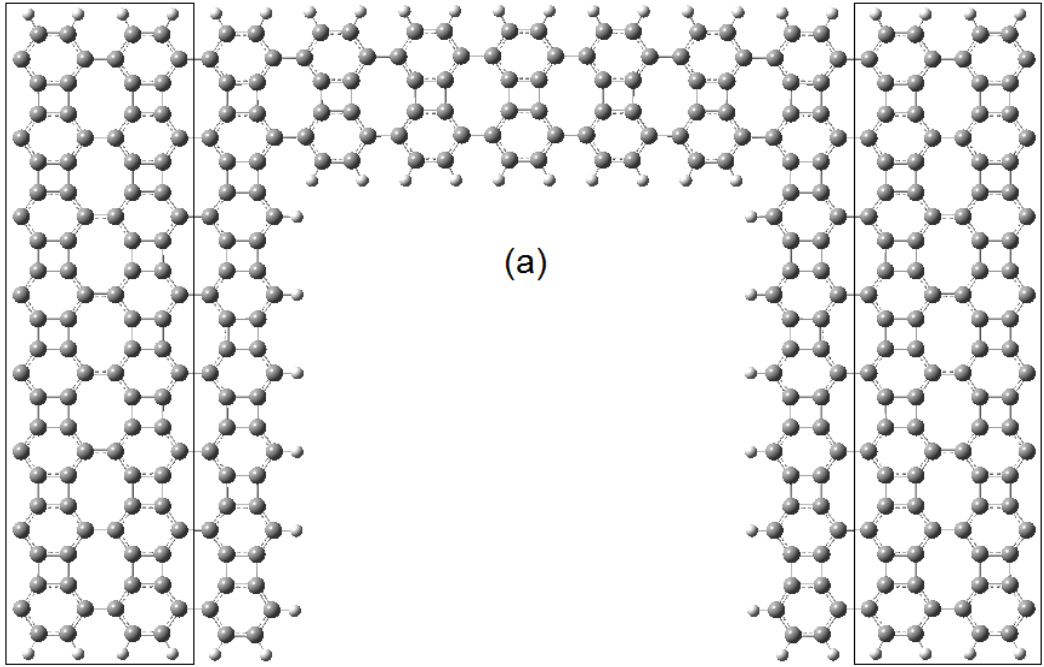

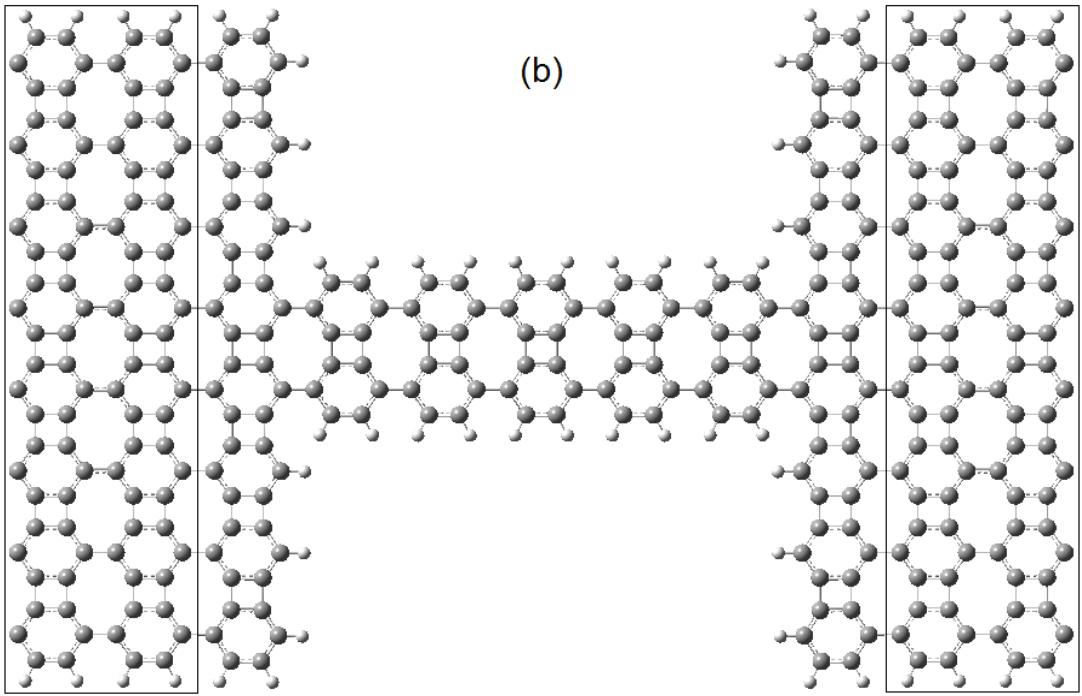

**Figure 2.** (a) Edge and (b) center models of electronic devices. Left and right electrodes are indicated by rectangles.

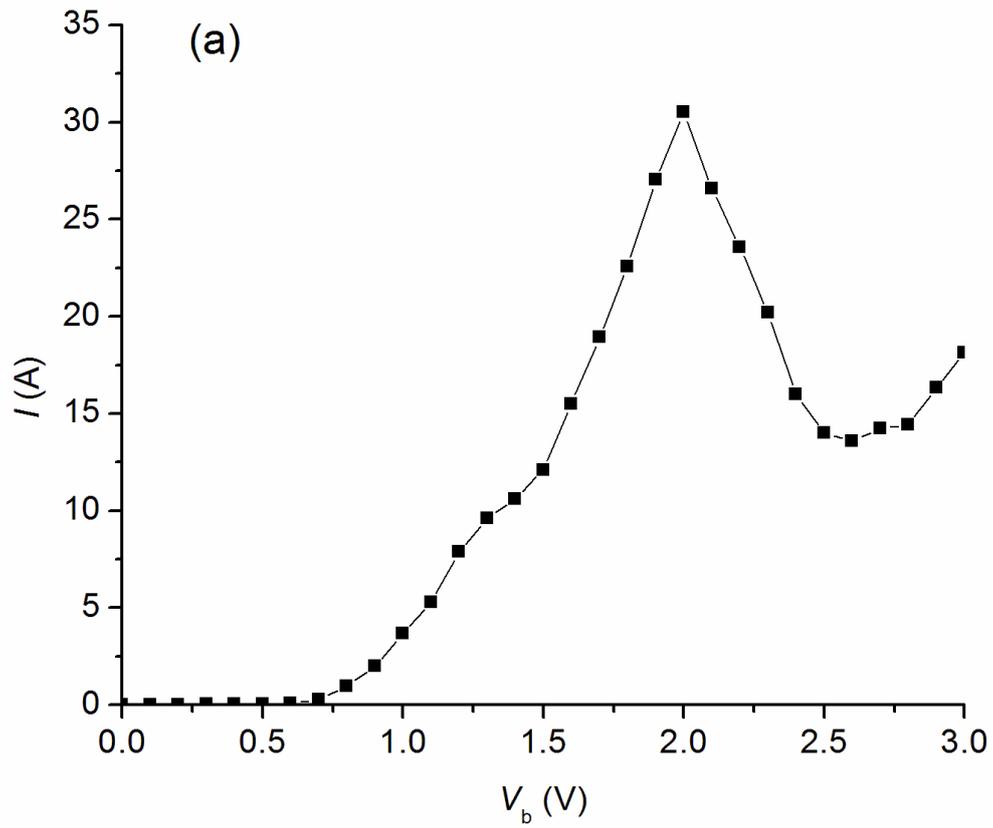
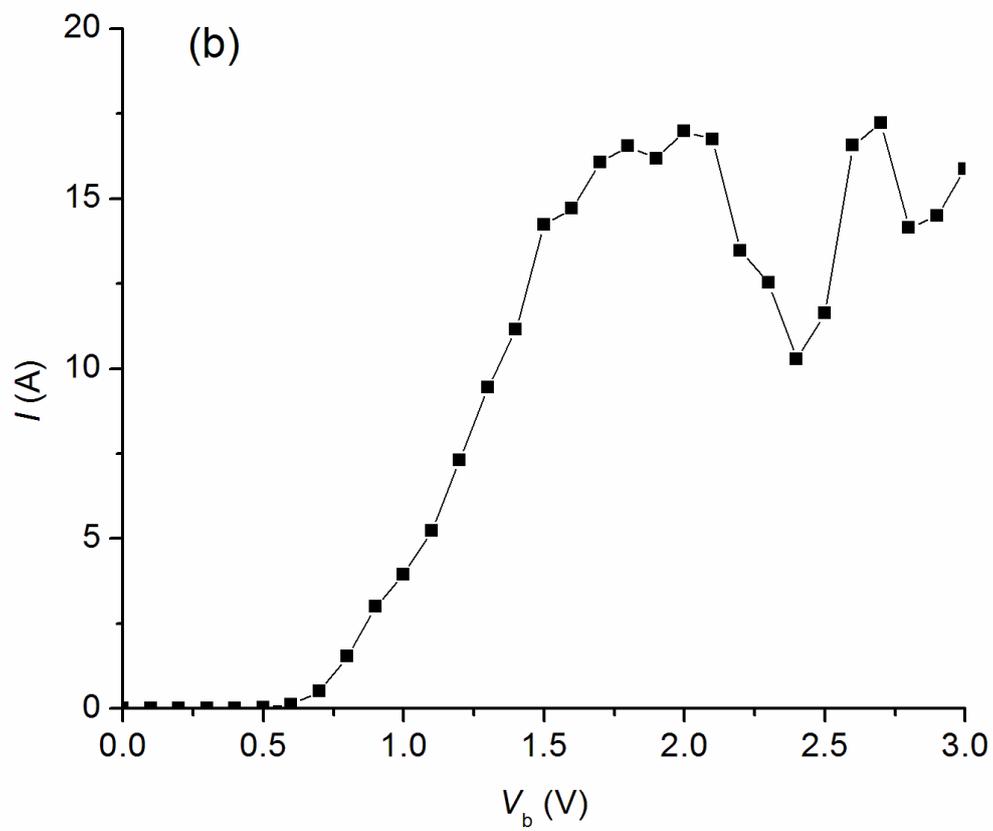

**Figure 3.** *I-V* characteristics of (a) the edge and (b) center models.

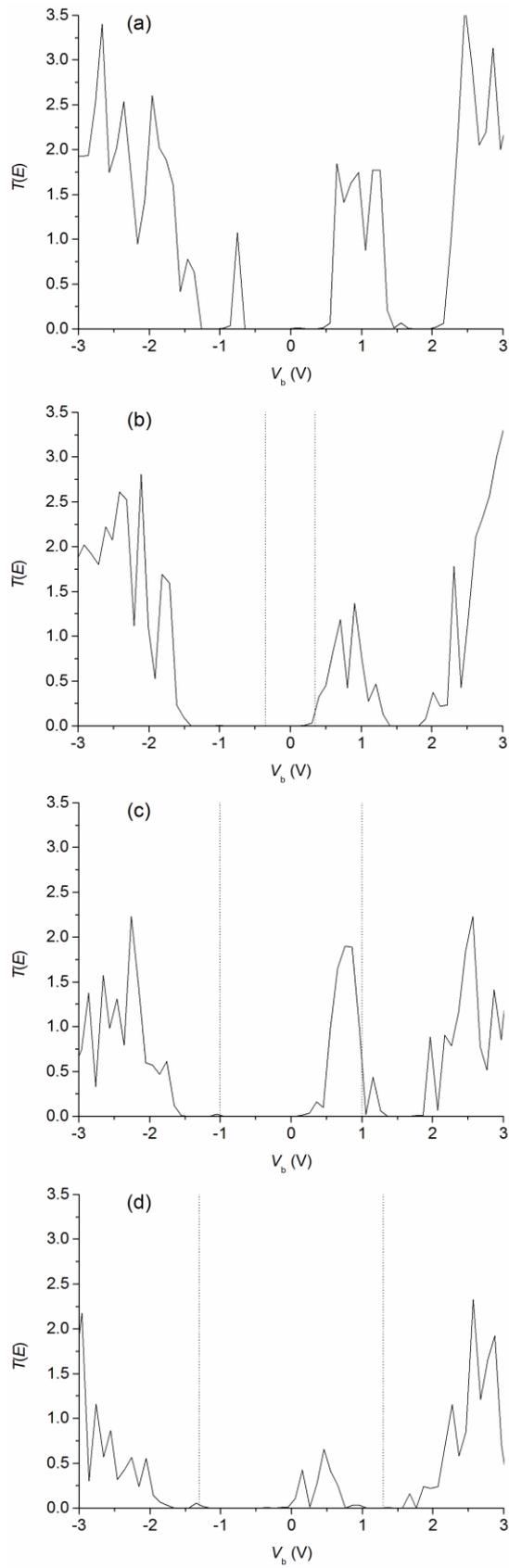

**Figure 4.** Transmission spectra of the edge model at bias of (a) 0, (b) 0.7, (c) 2.0 and (d) 2.6 V. Fermi levels are set to zero.